\documentclass[twocolumn,amsmath,showpacs,amsfonts,aps,prc,floatfix]{revtex4}
\usepackage{graphicx}
\usepackage{bm}

\begin{document}

\title{Fluctuating initial condition and smoothening effect on elliptic and triangular flow}
 
 \author{Md. Rihan Haque}
\email[E-mail:]{rihanphys@veccal.ernet.in}
\affiliation{Variable Energy Cyclotron Centre, 1/AF, Bidhan Nagar, 
Kolkata 700~064, India}
 \author{Victor Roy}
\email[E-mail:]{victor@veccal.ernet.in}
\affiliation{Variable Energy Cyclotron Centre, 1/AF, Bidhan Nagar, 
Kolkata 700~064, India}
\author{A. K. Chaudhuri}
\email[E-mail:]{akc@veccal.ernet.in}
\affiliation{Variable Energy Cyclotron Centre, 1/AF, Bidhan Nagar, 
Kolkata 700~064, India}

\begin{abstract}

In heavy ion collisions, event-by-event fluctuations in participating nucleon
positions can lead to triangular flow. Generally, one uses Monte-Carlo Glauber model to obtain the participating nucleon positions. To use in a hydrodynamic model, the positions needs to be smoothened. We study the effect of smoothening of Glauber Monte-Carlo initial conditions on elliptic and triangular flow. 
It is shown that
integrated as well as differential
elliptic and triangular flow remain largely unaltered, irrespective of functional form of the smoothening function, or the smoothening parameter
\end{abstract}


\date{\today}  

\maketitle


Azimuthal distribution of the produced particles is one of the important observables in ultra-relativistic nuclear collisions.  In a non-zero impact parameter collision between two identical nuclei, the collision zone is asymmetric. Multiple collisions transform the initial asymmetry   into momentum anisotropy. Momentum anisotropy is best studied by decomposing it   in a Fourier series, 
 
\begin{equation} \label{eq1}
\frac{dN}{d\phi}=\frac{N}{2\pi}\left [1+ 2\sum_n v_n cos(n\phi-n\psi_n)\right ], n=1,2,3...
\end{equation} 
 
\noindent   $\phi$ is the azimuthal angle of the detected particle and 
$\psi_n$ is the  plane of the symmetry of initial collision zone. For smooth initial matter distribution, plane of symmetry of the collision zone coincides with the reaction plane (the plane containing the impact parameter and the beam axis), 
$\psi_n \equiv \Psi_{RP}, \forall n$. The odd Fourier coefficients are zero by symmetry. However, fluctuations in the positions of the participating nucleons can lead to non-smooth density distribution, which will fluctuate on event-by-event basis.  
The participating nucleons then determine the symmetry plane ($\psi_{PP}$), which fluctuate around the reaction plane \cite{Manly:2005zy}. As a result odd harmonics, which were exactly zero for smoothed initial distribution, can be developed. It has been conjectured that third hadronic $v_3$, which is response of the initial triangularity of the medium, is responsible for the observed structures in two particle correlation in Au+Au collisions \cite{Mishra:2008dm},\cite{Mishra:2007tw},\cite{Takahashi:2009na},\cite{Alver:2010gr},\cite{Alver:2010dn},\cite{Teaney:2010vd}. The ridge structure in $p{\bar p}$ collisions also has a natural explanation if odd harmonic flow develops.  Recently, ALICE collaboration has observed odd harmonic flows    in Pb+Pb collisions \cite{:2011vk}. In most central collisions, the elliptic flow ($v_2$) and triangular flow ($v_3$) are of similar magnitude. In peripheral collisions however, elliptic flow dominates. 

In a hydrodynamic model, collective flow is a response of the spatial asymmetry of the initial state. For example, elliptic flow   is the response of  ellipticity of the initial medium, triangular flow is the response of the initial triangularity of the medium and so on. In theoretical simulations, one generally uses Monte-Carlo Glauber model to obtain the event-by-event initial conditions.
In a Monte-Carlo Glauber model, according to the density distribution of the colliding nuclei,    two nucleons are randomly chosen. If the transverse separation between them is below a certain distance, they are assumed to interact. Transverse position of the participating nucleons is then known in each event and will fluctuate from event-to-event. If a particular event has $N_{part}$ participants,  participants positions in the transverse plane can be labeled as, $(x_1,y_1), (x_2,y_2)....(x_{npart},y_{npart})$. The energy density in the transverse plane can be approximated as,

\begin{equation}\label{eq2}
\varepsilon(x,y) \propto \sum_{i=1}^{npart}  \delta(x-x_i,y-y_i)
\end{equation}

However, fluid dynamical model require continuous density distribution and discrete distribution as in Eq.\ref{eq2} cannot be evolved. 
To use in a hydrodynamic model, the discrete density distribution has to be converted into a smooth energy-density distribution. This can be done by smearing the discrete participating nucleon position by some smoothening function, $\delta(x-x_i,y-y_i) \rightarrow g(x-x_i,y-y_i,\zeta_1,\zeta_2..)$,
$\zeta_i$ being parameters of the smoothening function $g$. Incidentally, even though smoothening of the  nucleon positions is a must if one uses Monte-Carlo-Glauber model, existing literature in general do not elaborate on the procedure. However, if the flow coefficients are to be used for diagnostic purpose, e.g. to constrain the viscosity over entropy ratio, it is important to know the effect of smoothening. In the present brief report, we study the effect of smoothening of
Monte-Carlo Glauber model initial condition on the elliptic and triangular flow. It will be shown that   the flow coefficients depend minimally on the smoothening function.

In the following, we write the smoothen out energy density in the transverse plane as, 

\begin{equation} \label{eq3}
\varepsilon(x,y)=\varepsilon_0\sum_{i=1}^{npart}  g(x,y,x_i,y_i,\zeta_1,\zeta_2....)
\end{equation}

\noindent where $g(x-x_i,y-y_i,\zeta_1,\zeta_2....)$ 
is the smoothening function and $\zeta_i$'s are the parameters of the smoothening function. $\varepsilon_0$ is a parameter, which is fixed such that
event averaged particle multiplicity reproduces the experimental value.
In the following, we consider two smoothening functions, (i) a Woods-Saxon distribution and (ii) a Gaussian distribution. The Woods-Saxon distribution is,

\begin{eqnarray}
g_{WS}(x-x_i,y-y_i,C,a) &\propto& \frac{1}{1+e^{\frac{\sqrt{(x-x_i)^2+(y-y_i)^2}-C}{a} }} \label{eq4}
\end{eqnarray}

\begin{figure}[t]
 \center
 \resizebox{0.35\textwidth}{!}{%
  \includegraphics{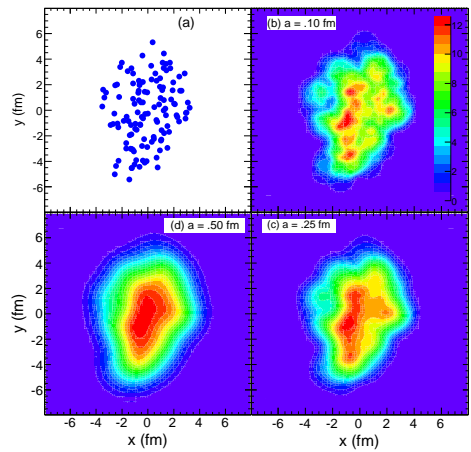}
}
\caption{(color online) (a) discrete positions of participating nucleons in the transverse plane in a typical event from Monte-Carlo Glauber model, in b=7 fm Au+Au collision. (b)-(d) density distribution obtained by smoothening the discrete position by Woods-Saxon function, with diffuseness parameter, $a$=0.1, 0.25 and 0.5 fm. $C=1 fm$.}
\label{F1}
\end{figure}

The half radius $C$ in Eq.\ref{eq4} is kept fixed $C=1 fm$. Only the   diffuseness parameter $a$ is varied to obtain different smoothening.  In the following, we consider four different values, a=0.05, 0.1, 0.25 and 0.5 fm.   The effect of smoothening on the energy density distribution can be seen in Fig.\ref{F1}. In panel (a), the transverse distribution of participating nucleons, in b=6.5 fm Au+Au collisions, for a typical Monte-Carlo event is shown. The discrete positions are smoothened with a Woods-Saxon distribution. In panel (b)-(d), the smoothened out energy density distribution, for diffuseness parameter $a$=0.1, 0.25 and 0.5 fm are shown. One notes that structures in the density profiles are diffused as the diffuseness increase. In the following, we study the effect of smoothening on elliptic and triangular flow. 

To show the dependence on the functional form, we have also smoothened the participant nucleons positions with a  Gaussian function,

\begin{eqnarray}
g_{gauss}(x-x_i,y-y_i,\sigma) &\propto& e^{-\frac{{(x-x_i)^2+(y-y_i)^2}}{2\sigma^2}  } \label{eq5}
\end{eqnarray}

Two values of the Gaussian width, $\sigma$=0.1 and 0.5 fm are considered.
In the following, we will refer to the diffuseness parameter ($a$)of the Woods-Saxon function and the width ($\sigma$) of the Gaussian   as the smoothening parameter ($s$).  

With fluctuating initial conditions, in each event, space-time evolution of the fluid is obtained by solving the energy-momentum conservation equations $\partial_\mu T^{\mu\nu}=0$,  in $(\tau=\sqrt{t^2-z^2},x,y,\eta_s=\frac{1}{2}\ln\frac{t+z}{t-z})$ coordinate system, with the code AZHYDRO-KOLKATA  \cite{Chaudhuri:2008sj}. We assume boost-invariance. We disregard any dissipative effect.
Ideal hydrodynamics equations are closed with an equation of state (EoS) $p=p(\varepsilon)$.
Currently, there is consensus that the confinement-deconfinement transition is a cross over and the cross over or the pseudo critical temperature for the  transition  is
$T_c\approx$170 MeV \cite{Aoki:2006we,Aoki:2009sc,Borsanyi:2010cj,Fodor:2010zz}.
In the present study, we use an equation of state where the Wuppertal-Budapest \cite{Aoki:2006we,Borsanyi:2010cj} 
lattice simulations for the deconfined phase is smoothly joined at $T=T_c=174$ MeV, with hadronic resonance gas EoS comprising all the resonances below mass $m_{res}$=2.5 GeV. Details of the EoS can be found in \cite{Roy:2011xt}.

   \begin{figure}[t]
 \center
 \resizebox{0.35\textwidth}{!}{%
  \includegraphics{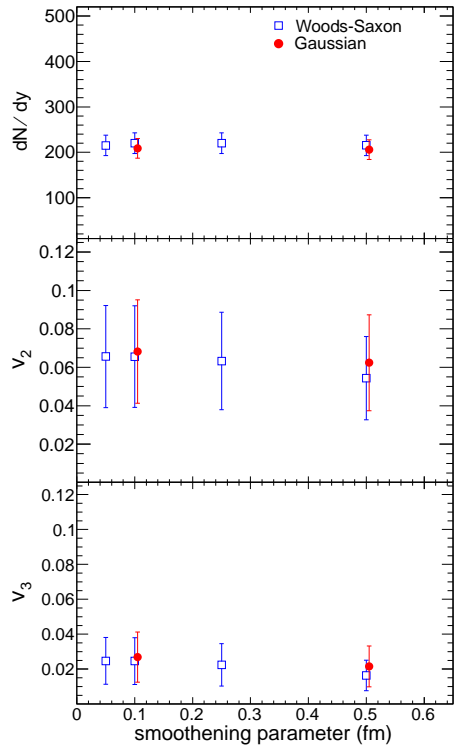}
}
\caption{(color online)  Charged particles multplicity (top panel), integrated elliptic flow (the middle panel) and integrated triangular flow (the bottom panel), as function of smoothening parameter (s). The filled and solid symbols are for Woods-Saxon and Gaussian smoothening function respectively.}
\label{F2}
\end{figure} 

In addition to the initial energy density for which we use the smoothened out Monte-Carlo Glauber model,  solution of hydrodynamic  equations   requires to specify the thermalisation or the initial time $\tau_i$ and fluid velocity ($v_x(x,y),v_y(x,y)$).  A freeze-out prescription is also needed to convert the information about fluid energy density and velocity to invariant particle distribution.  We assume that the fluid is thermalized at $\tau_i$=0.6 fm and the initial fluid velocity is zero, $v_x(x,y)=v_y(x,y)=0$. The freeze-out is fixed at $T_F$=130 MeV. 
We use Cooper-Frye formalism to obtain the invariant particle distribution from the freeze-out surface. From the  invariant distribution, harmonic flow coefficients are obtained as \cite{arXiv:1104.0650},

    \begin{figure}[t]
 \vspace{0.3cm} 
 \center
 \resizebox{0.35\textwidth}{!}{%
  \includegraphics{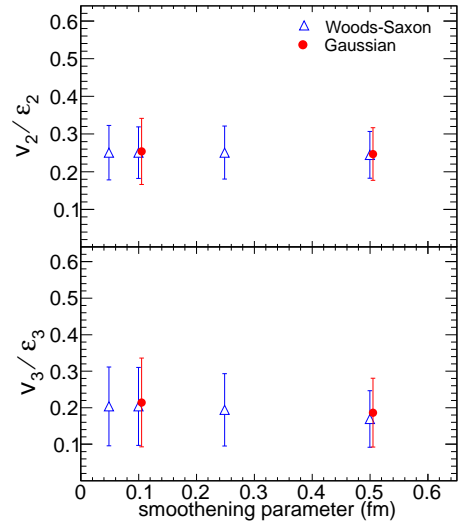}
}
\caption{(color online) Top panel: integrated elliptic flow scaled by initial eccentricity as a function of smoothening parameter (s). The filled and solid symbols are for Woods-Saxon and Gaussian smoothening function respectively. Bottom panel: same for the initial triangularity scaled integrated triangular flow. }
\label{F3}
\end{figure}  

\begin{eqnarray}
v_n(y,p_T)e^{in\psi_n(y,p_T)}&=&\frac{\int d\phi e^{in\phi} \frac{dN}{dy  p_Tdp_T d\phi}}  {\frac{dN}{dy p_Tdp_T}} \label{eq6}\\
  v_n(y)e^{in\psi_n(y)}&=& \frac{ \int p_T dp_T d\phi e^{in\phi} \frac{dN}{dy p_T dp_T d\phi} } { \frac{dN}{dy} } \label{eq7}
\end{eqnarray}
  
In a boost-invariant version of hydrodynamics, flow coefficients are rapidity independent and in the following.  
Present simulations are suitable only for central rapidity, $y\approx$0, where boost-invariance is most justified. Hereafter, we drop the rapidity dependence. $\psi_n$ in Eq.\ref{eq6},\ref{eq7} is the reaction plane angle. 
We use the following definitions of initial eccentricity $\epsilon_2$ and triangularity $\epsilon_3$ \cite{Alver:2010gr},\cite{Alver:2010dn},\cite{Teaney:2010vd},

\begin{eqnarray} 
\epsilon_2 e^{i2\psi_2} &=&-\frac{\int \int \varepsilon(x,y) r^2 e^{i2\phi}dxdy}{\int \int\varepsilon(x,y) r^2 dxdy} \label{eq8}\\
\epsilon_3 e^{i3\psi_3} &=&-\frac{\int \int \varepsilon(x,y) r^3 e^{i3\phi}dxdy}{\int \int\varepsilon(x,y) r^3 dxdy} \label{eq9}
\end{eqnarray}

\begin{figure}[t]
 \center
 \resizebox{0.35\textwidth}{!}{%
  \includegraphics{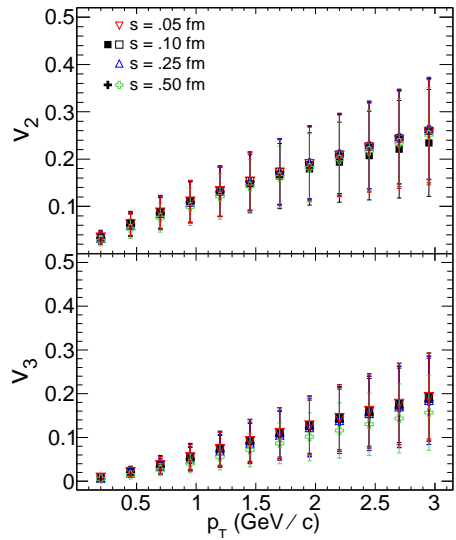}
}
\caption{(color online) Dependence of differential elliptic (the top panel) and  
  differential triangular  (the bottom panel) flow, on the smoothening parameter (s). The open and filled symbols are respectively for Woods-Saxon and  Gaussian   smoothening function.}
\label{F4}
\end{figure}

In the following we show simulation results for 30-35\%  Au+Au collisions at $\sqrt{s}_{NN}$=200 GeV. We have considered $N_{event}=100$ events.  Recently in \cite{Chaudhuri:2011qm}, it was shown that with fluctuating initial conditions, event averaged as well as variance of elliptic flow and triangular flow remain approximately unaltered for $N_{event}$=50-2500.  $N_{event}=100$ is then sufficiently large to study the effect of smoothening on flow coefficients.
The parameter $\varepsilon_0$ in Eq.\ref{eq3}, is fixed such that irrespective of the smoothening function or smoothening parameter,     hydrodynamic evolution approximately reproduces the experimental charged particles multiplicity. Results of our simulations for $N=100$ Monte-Carlo Glauber model events are shown in Fig.\ref{F2}-\ref{F4}.  The filled/open symbols are event averaged values and the   bars represent the variance. The filled symbols are obtained with smoothening with the Woods-Saxon  function, the open symbols are obtained with the Gaussian function. The top panel of Fig.\ref{F2} shows that the simulations are constrained such that the average charged particles multiplicity is $dN/dy\approx 210$, irrespective of the smoothening function and the smoothening parameter. In the middle panel of Fig.\ref{F2}, dependence of the integrated elliptic flow  on the smoothening
function and smoothening  parameter is shown. Both for Gaussian and Woods-Saxon smoothening function, elliptic flow do not show any dependence on the smoothening parameter. Note that for the Woods-Saxon function, the smoothening parameter is varied by a factor of 10 and
for the Gaussian function, by a factor of 5.  Even though the smoothening parameter is changed a factor of 5-10, event averaged $v_2$ and their variances remain approximately unchanged.  More interestingly,  smoothening with Gauss or Woods-Saxon function, both produces approximately same $v_2$. 

In the botom panel of Fig.\ref{F2}, simulation results for the triangular flow is shown. 
As it was observed for the elliptic flow, integrated triangular flow also do not show any   dependence on the functional form of the smoothing function or on the smoothening parameter. Both for Gaussian function and Woods-Saxon function, approximately similar value is obtained for the triangular flow. Event averaged triangular flow also show marginal dependence on the smoothening parameter. 
In event-by-event hydrodynamics, with Monte-Carlo Glauber model initial conditions, the elliptic and triangular flow seems to be independent of the smoothening function as well as of the smoothening parameter.

    
In ideal hydrodynamics, anisotropic flow is expected to scale with   initial asymmetry of the reaction, quantified in terms of initial eccentricity and initial triangularity etc. Presently, we donot study the scaling behavior of flow coefficient. Rather, we study the effect of smoothening on the scaled observables,   
the initial eccentricity scaled elliptic flow ($v_2/\epsilon_2$) and initial triangularity scaled triangular flow ($v_2/\epsilon_3$). The simulation results are shown in Fig.\ref{F3}.    As it was for the elliptic and triangular flow, initial eccentricity scaled elliptic flow as well as initial triangularity scaled triangular flow also remain unaltered for Gaussian and Woods-Saxon smoothening functions. It also show little dependence on the smoothening parameter. 
 
Lastly, in Fig.\ref{F4}, 
effect of smoothening on the   differential elliptic and triangular flow is shown. As before, the filled/open symbols are for Woods-Saxon/Gaussian smoothening functions. As it was for the integrated $v_2$, differential $v_2$ also depend marginally on the smoothening function or on the smoothening parameter. For Gaussian 
smoothening function, event averaged $v_2(p_T\approx 1.5 GeV)$ changes by $\sim$ 3\% for a factor of 5 change in the smoothening parameter, from 0.1 to 0.5. The change is even less for the Woods-Saxon function. Indeed, within the uncertainties, the differential $v_2$ is approximately independent of the smoothening function and of smoothening parameter. The differential 
triangular flow is also remains approximately independent of the smoothening function and smoothening parameter.  However, absolute change is marginally more. For example, for Gaussian smoothening function, event averaged $v_2(p_T\approx 1.5 GeV)$ is reduced by $\sim$ 10\% for a factor of 5 increase in the smoothening parameter. The change is even less ($\sim$7\%) for a factor of 10 change in the smoothening parameter of the Woods-Saxon function.

To summarise, in event-by-event hydrodynamics, discrete participant nucleon position from Monte-Carlo Glauber model calculation required to be smoothened. 
We have investigated the effect smoothening on   elliptic and triangular flow. In each event, participating nucleons position from Monte-Carlo Glauber model calculations are smoothened with either a Gaussian function or a Woods-Saxon function. Smoothening is controlled by a smoothening parameter, e.g. diffuseness for Woods-Saxon function and width for the Gaussian function. It was shown that
integrated as well as differential
elliptic and triangular flow remain largely unaltered, irrespective of functional form of the smoothening function, or the smoothening parameter.  \\

\textbf{Acknowledgements}\\
Md. Rihan Haque is supported by the DAE-BRNS project grant No. 2010/21/15-BRNS/2026.

\end{document}